\documentclass[%
 12pt,
 reprint,
  twocolumn,
 showpacs,
 showkeys,
 preprintnumbers,
 amsmath,amssymb,
 aps,
  pra,
  longbibliography,
 ]{revtex4-1}

\usepackage[breaklinks=true,colorlinks=true,anchorcolor=blue,citecolor=blue,filecolor=blue,menucolor=blue,pagecolor=blue,urlcolor=blue,linkcolor=blue]{hyperref}
\usepackage{url}

\usepackage{graphicx}

\usepackage[usenames,dvipsnames]{xcolor}

\usepackage{musixtex}

\begin{document}

\title{Quantum music}


\author{Volkmar Putz}
\affiliation{P\"adagogische Hochschule Wien, Grenzackerstra\ss e 18, A-1100
    Vienna, Austria}
\email{volkmar.putz@phwien.ac.at}

\author{Karl Svozil}
\affiliation{Institute for Theoretical Physics, Vienna
    University of Technology, Wiedner Hauptstra\ss e 8-10/136, A-1040
    Vienna, Austria}
\email{svozil@tuwien.ac.at} \homepage[]{http://tph.tuwien.ac.at/~svozil}


\begin{abstract}
We consider ways of conceptualizing, rendering and perceiving quantum music, and quantum art in general. Thereby, we give particular emphasis to its non-classical aspects, such as coherent superposition and entanglement.
\end{abstract}

\maketitle

\section{Introduction}

As a {\it caveat} we would like to state upfront that we shall primarily deal with artistic {\em expressibility}
rather than with  aesthetics  --
we take it for granted that the human perception of art is invariably bound by the human neurophysiology and hence is subject to a rather narrow
bracket or ``aesthetic bandwidth'' in between monotony and chaos~\cite{svozil-aesthetic_complexity}.
One may speculate that art in the past centuries until today, from the {\it Belle {\'E}poque} onward,
is increasingly dominated by {\em scarcity} and the cost of creation and rendition.
Those forms of artistic expressions, such as architecture,
for which an increase of complexity, in particular ornamentation, are costly, tend to become more monotonous,
whereas in other artistic domains such as music the tendency to increase complexity by sacrificing harmony has encouraged compositions which are notoriously
difficult to perceive.

Moreover, human neurophysiology suggests that artistic beauty
cannot easily be disentangled from sexual attraction.
It is, for instance, very difficult to appreciate Sandro Botticelli's {\em Primavera}, the arguably ``most beautiful painting ever painted,''
when a beautiful woman or man is standing in front of that picture.
Indeed so strong may be the distraction, and so deep the emotional impact, that it might not be unreasonable to speculate
whether aesthetics,
in particular beauty and harmony in art, could be best understood in terms of surrogates for natural beauty.
This might be achieved through the process of artistic creation, idealization and ``condensation.''
In this line of thought, in Hegelian terms,
artistic beauty is the sublimation, idealization, completion, condensation and augmentation of natural beauty.

Very different from  Hegel who asserts that artistic beauty is
{\it ``born of the spirit and born again, and the higher the spirit and its productions are above nature and its phenomena,
the higher, too, is artistic beauty above the beauty of nature~\cite[Part I, Introduction]{Hegel-Aesthetics}''}
we believe that human neurophysiology can hardly be disregarded for the human creation and perception of art; and,
in particular, of beauty in art.
Stated differently, we are inclined to believe that humans are so invariably determined by (or at least intertwined with) their natural basis that any neglect of it
results in a humbling experience of irritation or even outright ugliness, no matter what social pressure groups
or secret services~\cite{Wilford-CIA}
may want to promote.

Thus, when it comes to the intensity of the experience, the human perception of
artistic beauty, as sublime and refined as it may be, can hardly transcend natural beauty in its full exposure.
For example, it is not unreasonable to suspect that
the {\em Taj Mahal} could never compensate its commissioner {\em Mughal emperor Shah Jahan} for the loss of his beloved third wife {\it Mumtaz Mahal}.
In that way, art represents both the capacity as well as the humbling ineptitude of its creators and audiences.

Let us leave these idealistic realms and come back to the quantization of musical systems.
The universe of music consists of an infinity -- indeed a continuum -- of tones and ways to compose, correlate and arrange them.
It is not evident how to quantize sounds, and in particular music, in general.
One way to proceed would be a microphysical one: to start with frequencies of sound waves in air
and quantize the spectral modes of these (longitudinal) vibrations very similar to phonons in solid state physics~\cite{FetterWale}.

For the sake of relating to music, however, we shall pursue a different approach that is not dissimilar
to the Deutsch-Turing approach to universal (quantum) computability~\cite{mermin-07},
or Moore's automata analogues
to complementarity~\cite{e-f-moore}:
we shall quantize a musical instrument; in particular, a piano.
To restrict our considerations even further we shall only be concerned with an octave,
realized by the eight white keyboard keys
typically written $c$, $d$, $e$, $f$, $g$, $a$, $b$, $c'$ (in the C major scale), respectively.
Of course, from a musical point of view, it would be preferable to use the entire chromatic scale;
unfortunately, this could increase the complexity of the argument without gaining conceptual advantages.

In analogy to quantum information we shall first consider the quantization of tones.
We shall introduce a nomenclature in analogy to classical musical representation.
Then we will introduce typical quantum mechanical features such as the coherent superposition of classically distinct tones,
as well as entanglement and complementarity in music.

\section{Quantum musical tones}

In what follows, we shall quantize musical instruments, in particular, a piano.
Thereby we have to make formal choices which are not unique.
We shall mention alternatives as we develop the theory.

We consider a quantum octave in the C major scale, which classically consists
of the tones  $c$, $d$, $e$, $f$, $g$, $a$, $b$, and $c'$,
represented by eight consecutive white keys on the piano. (Other scales are straightforward.)
At least three ways to quantize this situation can be given: (i) bundling octaves,
as well as considering pseudo-field theoretic models
treating notes as (ii) bosonic or (iii) fermionic field modes.

\subsection{Bundling octaves into single observables}

We could treat the seven tones
$c$, $d$, $e$, $f$, $g$, $a$, and $b$ as belonging to disjoint events
(maybe together with the null event $0$) whose probabilities should add up to unity.
This would essentially suggest a formalization by a seven (or eight) dimensional Hilbert space ${\mathbb C}^7$  or ${\mathbb C}^8$)
with the standard Euclidean scalar product.
This Hilbert space represents a full octave.
In the quantum piano case, different observables
correspond to different octave blocks (realized by different equal-dimensional Hilbert spaces)
on the keyboard.

\begin{figure}
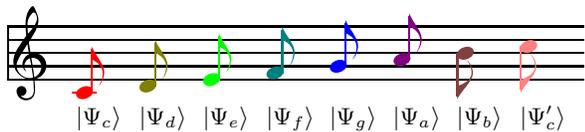

\begin{center}
\begin{music}
\startextract
\NOtes\zsong{$\vert \Psi_c \rangle$}{\color[rgb]{1,0,0}\ca{c}}\en
\NOtes\zsong{$\vert \Psi_d \rangle$}{\color[rgb]{0.5,0.5,0}\ca{d}}\en
\NOtes\zsong{$\vert \Psi_e \rangle$}{\color[rgb]{0,1,0}\ca{e}}\en
\NOtes\zsong{$\vert \Psi_f \rangle$}{\color[rgb]{0,0.5,0.5}\ca{f}}\en
\NOtes\zsong{$\vert \Psi_g \rangle$}{\color[rgb]{0,0,1}\ca{g}}\en
\NOtes\zsong{$\vert \Psi_a \rangle$}{\color[rgb]{0.5,0,0.5}\ca{h}}\en
\NOtes\zsong{$\vert \Psi_b \rangle$}{\color[rgb]{0.5,0.25,0.25}\ca{i}}\en
\NOtes\zsong{$\vert \Psi_c' \rangle$}{\color[rgb]{1,0.5,0.5}\ca{j}}\en
\zendextract
\end{music}
\end{center}
\caption{(Color online) Temporal succession of quantum tones
$\vert \Psi_c \rangle$,
$\vert \Psi_d \rangle$,~$\ldots$,
$\vert \Psi_b \rangle$
in the C major scale
forming the octave basis ${\mathfrak B}$.}
\label{2015-qmusic-fig1}
\end{figure}

From now on we shall only consider the seven-dimensional case ${\mathbb C}^7$.
The seven tones forming one octave can then be represented as a basis ${\mathfrak B}$
of  ${\mathbb C}^7$ by forming the set theoretical union of the orthogonal
unit basis vectors; that is, ${\mathfrak B} = \{ \vert \Psi_c \rangle , \vert \Psi_d \rangle,\ldots \vert \Psi_b \rangle \}$,
where the basis elements are the Cartesian basis tuples
$\vert \Psi_c \rangle =(0,0,0,0,0,0,1)$,
$\vert \Psi_d \rangle =(0,0,0,0,0,1,0)$,~$\ldots$,
$\vert \Psi_b \rangle =(1,0,0,0,0,0,0)$
of ${\mathbb C}^7$.
Fig.~\ref{2015-qmusic-fig1} depicts the basis ${\mathfrak B}$ by its elements, drawn in different colors.

Then pure quantum musical states could be represented as unit vectors
$\vert \psi \rangle \in {\mathbb C}^7$
which are linear combinations of the basis ${\mathfrak B}$; that is,
\begin{equation}
\vert \psi \rangle =
\alpha_c \vert \Psi_c \rangle
+ \alpha_d \vert \Psi_d \rangle
+\cdots
+\alpha_b \vert \Psi_b \rangle
,
\label{2015-qmusic-e1}
\end{equation}
with coefficients $\alpha_i$
satisfying
$
\vert \alpha_c  \vert^2
+
\vert \alpha_d  \vert^2
+ \cdots  +
\vert \alpha_b  \vert^2
=1
$.
Equivalent representations of $\vert \psi \rangle$ are in terms of
the
one-dimensional subspace
$\{ \vert \phi \rangle  \mid \vert \phi \rangle = \alpha \vert \psi \rangle ,\, \alpha \in {\mathbb C}\}$
 spanned by $\vert \psi \rangle$, or by the projector
$\textsf{\textbf{E}}_\psi =  \vert \psi \rangle  \langle \psi  \vert$.

In most general terms (at least for this octave),  a musical composition
-- the succession of quantized tones as time goes by and the system evolves --
such as a melody,
would be obtained by the unitary permutation of the state $\vert \psi \rangle$.
The realm of such compositions would be spanned by the succession of all
unitary transformations $\textsf{\textbf{U}}: {\mathfrak B} \mapsto {\mathfrak B}'$
mapping some orthonormal basis ${\mathfrak B}$ into another orthonormal basis ${\mathfrak B}'$;
that is~\cite{Schwinger.60}, $\textsf{\textbf{U}} =   \sum_i \vert {\Psi'}_i \rangle  \langle \Psi_i  \vert$.

\subsection{Quantum musical parallelism}

If a classical auditorium listens to the quantum musical state  $\vert \psi \rangle $ in Eq.~\ref{2015-qmusic-e1},
then the individual  listeners  may perceive $\vert \psi \rangle $ very differently;
that is, they will hear only a {\em single one} of the different tones with probabilities  $
\vert \alpha_c  \vert^2
$, $
\vert \alpha_d  \vert^2
$, $\ldots$, and $
\vert \alpha_b  \vert^2
$, respectively.

Pointedly stated, a truly quantum music  never renders a  unique listening experience --
it might not be uncommon for part of the audience to hear different manifestations of the quantum musical composition made up of all varieties of successions of tones.
For instance, one listener may hear Mozart's {\it A Little Night Music, K~525}, whereas another listener
Prokoviev's {\it Le pas d'acier, Op~41}, and a third one would enjoy a theme from Marx's {\it Autumn Symphony (1921)}.
We could perceive this as quantum parallel musical rendition -- a classical audience may perceive one
and the same quantum musical composition very differently.

\begin{figure}
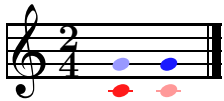

\begin{center}
\begin{music}
\generalmeter{\meterfrac24}
\startextract 
\Notes
{\color[rgb]{1,0.1,0.1}\zq c}{\color[rgb]{0.6,0.6,1}\zq g}  \enotes
\Notes
{\color[rgb]{1,0.6,0.6}\zq c}{\color[rgb]{0.1,0.1,1}\zq g}  \enotes
\Endpiece
\zendextract 
\end{music}
\end{center}
\caption{(Color online) A two-note quantum musical composition -- a natural fifth.}
\label{2015-qmusic-fig1a}
\end{figure}

For the sake of a demonstration,
let us try a two-note quantum composition.
We start with a pure quantum mechanical state
in the two-dimensional subspace spanned by  $\vert \Psi_c \rangle $ and $\vert \Psi_g \rangle$,
specified by
\begin{equation}
\vert \psi_1\rangle =
\frac45 \vert \Psi_c \rangle
+ \frac35 \vert \Psi_g \rangle = \frac15 \begin{pmatrix} 4 \\ 3  \end{pmatrix}
.
\end{equation}
$\vert \psi_1 \rangle$ would be detected by the listener as $c$ in 64\%
of all measurements (listenings), and as $g$ in 36\%
of all listenings.
Using the unitary transformation $\textsf{\textbf{X}}= \begin{pmatrix} 0 & 1 \\ 1 & 0 \end{pmatrix}$, the next quantum tone would be
\begin{equation}
\vert \psi_2 \rangle = \textsf{\textbf{X}}  \vert \psi_1 \rangle =
\frac35 \vert \Psi_c \rangle
+ \frac45 \vert \Psi_g \rangle = \frac15 \begin{pmatrix} 3 \\ 4  \end{pmatrix}.
\end{equation}
This means for the quantum melody of both quantum tones $\vert \psi_1\rangle$ and $\vert  \psi_2 \rangle$ in succession --
for score, see Fig.~\ref{2015-qmusic-fig1a} --
that in repeated measurements,
in $0.64^2 = 40.96\%$
of all cases $c-g$ is heard,
in $0.36^2 = 12.96\%$
of all cases $g-c$,
in $0.64\cdot0.36 = 23.04\%$
of all cases $c-c$ or $g-g$, respectively.
Thereby one single quantum composition can manifest
itself during listening in very different ways.

This offers possibilities of aleatorics in music far beyond the classical aleatoric methods of John Cage and his allies.

\subsection{Bose and Fermi model of tones}

An alternative quantization of music to the one discussed above
is in analogy to some fermionic or bosonic -- such as the electromagnetic -- field.
Just as the latter one in quantum optics~\cite{glauber:70,glauber-collected-cat}
and quantum field theory \cite{Weinberg-search}
is quantized by interpreting every single mode
(determined, for the electromagnetic field  for instance by
a particular frequency and polarization)
as a sort of ``container''
-- that is,
by allowing the occupancy of that mode to be either empty or any positive integer (and a coherent superposition thereof)
-- we obtain a vast realm of new musical expressions which cannot be understood in classical terms.

In what follows we shall restrict ourselves to a sort of ``fermionic field model'' of music
which is characterized by a binary, dichotomic situation, in which every tone has either null
or one occupancy, represented by $\vert 0 \rangle= (0,1)$ or $\vert 1 \rangle = (1,0) $, respectively.
Thus every state of such a tone can thus be formally represented by entities
of a two-dimensional Hilbert space ${\mathbb C}^2$, with the Cartesian standard basis
${\mathfrak B} =
\{
\vert 0 \rangle, \vert 1 \rangle
\}$.

Any note $\vert \Psi_i \rangle$ of the octave consisting of $\vert \Psi_c \rangle$,
$\vert \Psi_d \rangle$,~$\ldots$,
$\vert \Psi_b \rangle$,
$\vert \Psi_{c'} \rangle$
in the C major scale
can be represented by the coherent superposition of its null and one occupancies; that is,
\begin{equation}
\vert \Psi_i \rangle =
\alpha_i \vert 0_i \rangle
+
\beta_i \vert 1_i \rangle
,
\end{equation}
with
$
\vert \alpha_i \vert^2 + \vert \beta_i \vert^2 =1
$, $\alpha_i. \beta_i \in {\mathbb C}$.

\begin{figure}
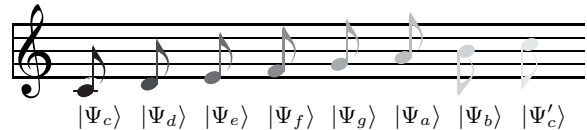

\begin{center}
\begin{music}
\startextract
\NOtes\zsong{$$}{\color{Black!0}\wh{d}}\en
\hspace{-10 mm}
\NOtes\zsong{$\vert \Psi_c \rangle$}{\color{Black!100}\ca{c}}\en
\NOtes\zsong{$\vert \Psi_d \rangle$}{\color{Black!85}\ca{d}}\en
\NOtes\zsong{$\vert \Psi_e \rangle$}{\color{Black!70}\ca{e}}\en
\NOtes\zsong{$\vert \Psi_f \rangle$}{\color{Black!55}\ca{f}}\en
\NOtes\zsong{$\vert \Psi_g \rangle$}{\color{Black!40}\ca{g}}\en
\NOtes\zsong{$\vert \Psi_a \rangle$}{\color{Black!30}\ca{h}}\en
\NOtes\zsong{$\vert \Psi_b \rangle$}{\color{Black!20}\ca{i}}\en
\NOtes\zsong{$\vert \Psi_c' \rangle$}{\color{Black!10}\ca{j}}\en
\zendextract
\end{music}
\end{center}
\caption{Temporal succession of tones $\vert \Psi_c \rangle$,
$\vert \Psi_d \rangle$,~$\ldots$,
$\vert \Psi_b \rangle$ in an octave in the C major scale with dicreasing mean occupancy.}
\label{2015-qmusic-fig2}
\end{figure}

At this stage the most important feature to notice is that every tone is characterized by the two coefficients
$\alpha$ and $\beta$, which in turn can be represented (like all quantized two-dimensional systems)
by a Bloch sphere, with two angular parameters.
If we restrict our attention (somewhat superficially) to real Hilbert space ${\mathbb R}^2$,
then the unit circle, and thus a single angle $\varphi$,
suffices for a characterization of the coefficients $\alpha$ and $\beta$.
Furthermore we may very compactly notate the mean occupancy of the notes by gray levels.
Fig.~\ref{2015-qmusic-fig2} depicts a sequence of tones in an octave in the C major scale with decreasing occupancy,
indicated as gray levels.

In this case, any non-monotonous unitary quantum musical evolution would have to involve the interaction of different tones;
that is, in the piano setting, across several keys of the keyboard. We shall come back to this later.

\section{Quantum musical coherent superposition}

\begin{figure}
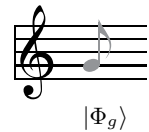

\begin{center}
\begin{music}
\startextract
\NOtes\zsong{$\vert \Phi_g \rangle$}{\color{Black!50}\ca{g}}\en
\zendextract
\end{music}
\end{center}
\caption{Representation of a 50:50 quantum tone $\vert \Phi_g \rangle =
\frac{1}{\sqrt{2}}
\left(
\vert 0_g \rangle
-
\vert 1_g \rangle
\right)
$ in gray (without indicating phase factors).}
\label{2015-qmusic-fig3}
\end{figure}

One of the mind-boggling quantum features is the possibility of the simultaneous formal ``existence'' of classically excluding musical states,
such as a 50:50 quantum $g$  in the C major scale obtained by sending $\vert 0_g \rangle $ through the Hadamard gate
$\textsf{\textbf{H}}= \frac{1}{\sqrt{2}}\begin{pmatrix} 1 & 1 \\ 1 & -1 \end{pmatrix}$,
resulting in
$
\frac{1}{\sqrt{2}}
\left(
\vert 0_g \rangle
-
\vert 1_g \rangle
\right)
$, and depicted in Fig.~\ref{2015-qmusic-fig3} by a 50 white 50 black; that is, gray, tone (though without the relative ``$-$'' phase).

In music, such experience of ``floating in as well as out of'' a tone -- faintly resembling the memory of having heard or not heard a particular tone or melody --
may not be totally foreign to audiences.
This new form of musical expression might contribute to novel musical experiences; in particular, if
any such coherent superposition can be perceived by the audience.
Note, however, that any attempt to ``amplify'' a coherent signal may be
in vain due to the inevitable introduction of noise~\cite{glauber,glauber-collected-cat}.

Schr\"odinger, in particular, was concerned about any such quantum coherence. When it is extended into macroscopic situations
it yields his cat paradox~\cite{schrodinger}; or his polemic regarding the ``jellification'' of the universe without measurement~\cite{schroedinger-interpretation}.
The puzzling basis of such alleged paradoxes is the seemingly impossibility of any conscious macroscopic individual entity to simultaneously pass through the two
slits of a double slit experiment; a property well verified for individual quanta~\cite{zeilinger:ds}.
From a purely formal point of view, any mixture of the two musical states amounts merely to a basis transformation in two-dimensional musical Hilbert space --
in this sense, the piano ``tuned to'' produce $\vert 0_g \rangle$ and $\vert 1_g \rangle$ needs to be ``retuned'' to
$\vert 0' \rangle =
\frac{1}{\sqrt{2}}
\left(
\vert 0_g \rangle
+
\vert 1_g \rangle
\right)
$
and
$\vert 1' \rangle =
\frac{1}{\sqrt{2}}
\left(
\vert 0_g \rangle
-
\vert 1_g \rangle
\right)
$, respectively.

\section{Quantum musical entanglement}

Quantum entanglement~\cite{schrodinger} is the property of multipartite quantum systems
to code information ``across quanta'' in such a way that the state of any individual quantum
is irreducibly indeterminate;
that is, not determined by the entangled multipartite state~\cite{zeil-99,zeil-Zuk-bruk-01}.
In other words, the entangled whole cannot be composed of its parts;
more formally, the composite state cannot be expressed as a product of states of the individual quanta.

\begin{figure}
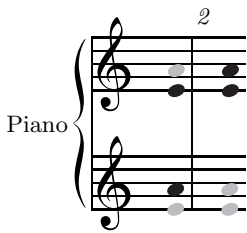

\begin{center}
\begin{music}
\parindent10mm
\instrumentnumber{1} 
\setname1{Piano} 
\setstaffs1{2} 
\startextract 
\notes
\hspace{1 mm}{\color{Black!30}\zq e}{\color{Black!100}\zq h}
|
\hspace{0 mm}{\color{Black!100}\zq e}{\color{Black!30}\zq h}
\en
\bar
\notes
\hspace{1 mm}{\color{Black!30}\zq e}{\color{Black!30}\zq h}
|
\hspace{0 mm}{\color{Black!100}\zq e}{\color{Black!100}\zq h}
\en
\zendextract 
\end{music}
\end{center}
\caption{Quantum musical entangled states
$\vert \Psi_{ea}^- \rangle$ and $\vert \Psi_{ea}^+ \rangle$
in the first bar,
and
$\vert \Phi_{ea}^- \rangle$ and $\vert \Phi_{ea}^+ \rangle$
und the second bar
(without relative phases).}
\label{2015-qmusic-fig4}
\end{figure}

A typical example of an entangled state is the
{\em Bell state},
$\vert \Psi^- \rangle$
or, more generally, states in the Bell basis spanned by the quantized notes $e$ and $a$; that is
\begin{equation}
\begin{split}
\vert \Psi^\pm \rangle = \frac{1}{\sqrt{2}}\left(\vert 0_e \rangle \vert 1_a \rangle \pm \vert 1_e \rangle \vert 0_a \rangle  \right),\\
\vert \Phi^\pm \rangle = \frac{1}{\sqrt{2}}\left(\vert 0_e \rangle \vert 0_a \rangle \pm \vert 1_e \rangle \vert 1_a \rangle  \right),\\
\end{split}
\label{2014-m-ch-fdvs-bellbasis}
\end{equation}
Indeed, a short calculation~\cite[Sec.~1.5]{mermin-07}
demonstrates that a necessary and sufficient condition
for entanglement among the quantized notes $e$ and $a$ is that the coefficients
$\alpha_1$,
$\alpha_2$,
$\alpha_3$,
$\alpha_4$
of their general composite state
$
\vert \Psi_{ga} \rangle =
\alpha_1 \vert 0_e \rangle \vert 0_a \rangle  +
\alpha_2 \vert 0_e \rangle \vert 1_a \rangle  +
\alpha_3 \vert 1_e \rangle \vert 0_a \rangle  +
\alpha_4 \vert 1_e \rangle \vert 1_a \rangle
$
obey $\alpha_1 \alpha_4 \neq \alpha_2 \alpha_3$.
This is clearly satisfied by Eqs.~(\ref{2014-m-ch-fdvs-bellbasis}).
Fig.~\ref{2015-qmusic-fig4} depicts the entangled music Bell states.

\begin{figure}
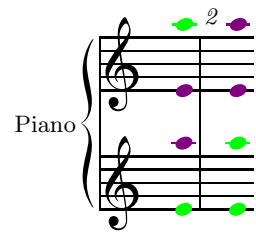

\begin{center}
\begin{music}
\parindent10mm
\instrumentnumber{1} 
\setname1{Piano} 
\setstaffs1{2} 
\startextract 
\notes
\hspace{1 mm}{\color[rgb]{0,1,0}\zq e}{\color[rgb]{0.5,0,0.5}\zq o}
|
\hspace{0 mm}{\color[rgb]{0.5,0,0.5}\zq e}{\color[rgb]{0,1,0}\zq o}
\en
\bar
\notes
\hspace{1 mm}{\color[rgb]{0,1,0}\zq e}{\color[rgb]{0,1,0}\zq o}
|
\hspace{0 mm}{\color[rgb]{0.5,0,0.5}\zq e}{\color[rgb]{0.5,0,0.5}\zq o}
\en
\zendextract 
\end{music}
\end{center}
\caption{(Color online) Quantum musical entangled states for bundled octaves
$\vert \Psi_{ea'}^- \rangle$ and $\vert \Psi_{ea'}^+ \rangle$
in the first bar,
and
$\vert \Phi_{ea'}^- \rangle$ and $\vert \Phi_{ea'}^+ \rangle$
in the second bar
(without relative phases).}
\label{2015-qmusic-fig5}
\end{figure}

We only remark that a very similar argument yields entanglement between different octaves.
Fig.~\ref{2015-qmusic-fig5} depicts this configuration for an entanglement between $e$ and $a'$.

\section{Quantum musical complementarity}

\begin{figure}
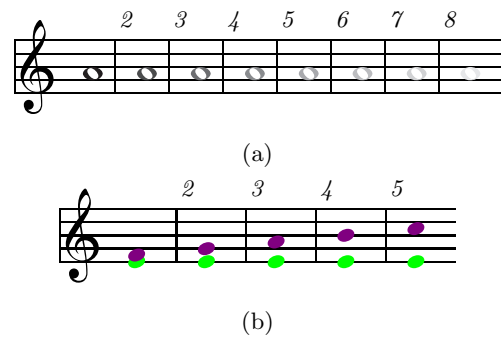

\begin{center}
\begin{music}
\startextract 
\notes
{\color{Black!100}\wh{h}}\enotes\bar
\notes {\color{Black!85}\wh{h}}\enotes\bar
\notes {\color{Black!70}\wh{h}}\enotes\bar
\notes {\color{Black!55}\wh{h}}\enotes\bar
\notes {\color{Black!40}\wh{h}}\enotes\bar
\notes {\color{Black!30}\wh{h}}\enotes\bar
\notes {\color{Black!20}\wh{h}}\enotes\bar
\notes {\color{Black!10}\wh{h}}
\enotes
\zendextract
\end{music}
(a)
\begin{music}
\startextract 
\Notes
{\color[rgb]{0,1,0}\zq e}{\color[rgb]{0.5,0,0.5}\zq f}  \enotes
\bar
\Notes
{\color[rgb]{0,1,0}\zq e}{\color[rgb]{0.5,0,0.5}\zq g}  \enotes
\bar
\Notes
{\color[rgb]{0,1,0}\zq e}{\color[rgb]{0.5,0,0.5}\zq h}  \enotes
\bar
\Notes
{\color[rgb]{0,1,0}\zq e}{\color[rgb]{0.5,0,0.5}\zq i}  \enotes
\bar
\Notes
{\color[rgb]{0,1,0}\zq e}{\color[rgb]{0.5,0,0.5}\zq j}  \enotes
\zendextract 
\end{music}
(b)
\end{center}
\caption{Temporal succession of complementary
tones (a) for binary occupancy $\vert \phi_a \rangle = \alpha_a \vert 0_a \rangle + \beta_a \vert 1_a \rangle$,
with $\vert \alpha_a \vert^2  + \vert \beta_a \vert^2  =1$ with increasing $\vert \alpha_a \vert$ (decreasing occupancy),
(b) in the bundled octave model,
separated by bars.}
\label{2015-qmusic-fig6}
\end{figure}

Although complementarity~\cite{pauli:1933} is mainly discussed in the context of observables,
we can present it in the state formalism by observing that,
as mentioned earlier, any pure state
 $\vert \psi \rangle$ corresponds -- that is, is in one-to-one correspondence (up to phase a factor) -- to the projector
$\textsf{\textbf{E}}_\psi =  \vert \psi \rangle  \langle \psi  \vert$.
In this way,
any two non-vanishing non-orthogonal and non-collinear states
$\vert \psi \rangle$
and
$\vert \phi \rangle$
with $0< \langle \phi \vert \psi \rangle < 1$
are complementary.
For the dichotomic field approach, Fig.~\ref{2015-qmusic-fig6} represents a configuration
of mutually complementary quantum tones for the note $a$ in the C major scale.

\section{Summary}

We have proposed the basic ideas  for a new kind of (quantum) music
by presenting a straightforward quantization of music, obtained by quantizing the ``white'' notes of a piano octave.
In this approach, generalizations to more than one octave, to the chromatic scale,
as well as to other musical instruments, appear to be straightforward.

We have also studied some non-classical features available to quantum music, such as coherent superposition
of classically distinct tones, tonal entanglement and complementarity.

We have pursued a strictly non-artistic, non-aesthetic approach.
In doing so we have merely attempted to extend music to the quantum realm.
No claims have been made that this realm is useful or necessary for aesthetics, or for musical expression.

One way to make use of this formalism is to get inspired by its freedom and new capacities;
even for quasi-classical analogues.

\begin{acknowledgments}
This research has been partly supported by FP7-PEOPLE-2010-IRSES-269151-RANPHYS.
\end{acknowledgments}


%

\end{document}